\documentclass[conference]{IEEEtran}
\IEEEoverridecommandlockouts
\usepackage{cite}
\usepackage{amsmath,amssymb,amsfonts}
\usepackage{algorithmic}
\usepackage{graphicx}
\usepackage{subcaption}
\usepackage{textcomp}
\usepackage{xurl}
\usepackage{xcolor}
\usepackage{pifont}
\usepackage[normalem]{ulem}
\newcommand{\cmark}{\ding{51}}%
\newcommand{\xmark}{\ding{55}}%
\def\BibTeX{{\rm B\kern-.05em{\sc i\kern-.025em b}\kern-.08em
    T\kern-.1667em\lower.7ex\hbox{E}\kern-.125emX}}
\begin{document}

\title{The Importance of Technical Distribution Network Limits in Dynamic Operating Envelopes

\thanks{This work has been supported in part by the European Union’s Horizon 2020 research and innovation programme under Grant Agreement No. 864298 (project ATTEST). The sole responsibility for the content of this document lies with the authors. It does not necessarily reflect the opinion of the Innovation and Networks Executive Agency (INEA) or the European Commission (EC). INEA or the EC are not responsible for any use that may be made of the information contained therein.

The work in this paper is exploratory and is not representative of GridQube's commercial offerings in this space.}
}

\author{\IEEEauthorblockN{Tomislav Antić}
\IEEEauthorblockA{\textit{Faculty of Electrical Engineering}\\\textit{and Computing} \\
\textit{University of Zagreb}\\
Zagreb, Croatia \\
tomislav.antic@fer.hr}
\and
\IEEEauthorblockN{Frederik Geth}
\IEEEauthorblockA{\textit{GridQube} \\
Springfield, Queensland, Australia \\
frederik.geth@gridqube.com}
\and
\IEEEauthorblockN{Tomislav Capuder}
\IEEEauthorblockA{\textit{Faculty of Electrical Engineering}\\\textit{and Computing} \\
\textit{University of Zagreb}\\
Zagreb, Croatia \\
tomislav.capuder\@fer.hr}
}

\maketitle

\begin{abstract}
End-users more often decide to invest in distributed generation (DG) units that help them in decreasing electricity bills and allow them to become a market player by selling the excess produced electricity. However, the installation of DG is often limited by technical constraints of the network, standards, and national grid codes. As a method for removing the mentioned obstacles, the potential of dynamic operating envelopes (DOEs) is recently becoming recognized as a way for maximizing the benefits of installing DG. In this paper, we present an improvement of the already developed models that often neglect voltage unbalance constraints or are not based on an optimization approach. To test the model, two realistic case studies are defined. The results show that not all technical constraints are equally important, that the voltage unbalance constraint impacts the calculated DOEs for single-phase installed DG units, and that neglecting the temporal and spatial component in determining the limitation power is  inadequate.
\end{abstract}

\begin{IEEEkeywords}
dynamic operating envelopes, low voltage networks, optimal power flow, technical constraints
\end{IEEEkeywords}

\section{Introduction}
Dynamic operating envelopes (DOEs) are a concept that has been proposed to allow distribution utilities to move away from static limits for distributed energy resources connecting to the distribution grid,  thereby enabling the energy transition by allowing more renewable generation \cite{fti2022}.
In Australia, there is growing momentum to enable flexible customer connections, with DOE calculation engines determining connection capacities in real-time \cite{CM2022}.
By freeing up additional network capacity, DOEs provide consumers with equitable access, improving on the inherently conservative nature of static export limits.
DOEs are typically envisioned to provide ranges of feasible import/export values -- \emph{not set points} -- at the customer connection level and are generally conceived to be physics-informed and explainable, with appropriate representation of technical limits of the system \cite{aer2022}.

Table~\ref{tab_lit_comparison} develops a comparison of key publications related to the development of DOE calculation engines, with a variety of approaches being taken. 
Certain papers focus on the case of active power exports only, however there seems to be a trend to consider both import and export limits for both active and reactive power.
The network model detail also varies between publications, with some of the papers using a balanced (a.k.a. positive sequence) model of the network impedances (size of the Z matrix is 1x1), others using sequence parameterization (positive, negative and zero sequence values are used to create 3x3 phase-coordinate impedance matrix) and others support a three-phase explicit-neutral representation (up to 4x4 as derived from the modified Carson's equations).
Papers represent a variety of network limits in power, voltage and/or current, however voltage unbalance limits are rarely discussed, despite them being part of standards of supply in public distribution networks. 

There are two main ways authors approach the calculation of the DOE values: either optimization-based or an iterative approach evaluating an existing power flow solver.
The optimization-based approaches are typically custom, being either proprietary or extending toolboxes such as \textsc{PowerModelsDistribution} \cite{FOBES2020106664} or \textit{pp OPF} \cite{ppopf}.
Iterative approaches \cite{ZABIHINIAGERDROODBARI2022119757,edge2022} can leverage commercial software to solve the power flows, which may also avoid conversion of network data formats. 
However, getting access to derivatives in commercial software is generally not supported, which makes it hard to build efficient optimization approaches that can assert the optimality conditions.

\begin{table*}[tb]
\renewcommand{\baselinestretch}{0.7}
\centering
\caption{Scope of different research outputs focusing on DOEs} \label{tab_lit_comparison}
{\footnotesize
\begin{tabular}{l c c c c  l } 
\hline
Reference &  Z size & P/Q? & import/export & optimization-based?     & Remarks \\ \hline
Blackhall, 2020 \cite{EVOLVE2020}   & 1x1 &  & & (agnostic)   & Design discussion \\ 
Petrou et al., 2020 \cite{9248975}  & 3x3  & P & both & \cmark   & Linearization\\ 
Liu B. \&  Braslavsky, 2020 \cite{Liu2022}  & 3x3 & P & both & \cmark  & Linearization, robust uncertainty \\ 
Liu M. et al., 2022 \cite{9816082} & 3x3 &  P+Q & both & \cmark    & Linearization  \\ %
Liu M. et al., 2021 \cite{9692618} & 3x3 & P & export & \cmark & Linearization\\ %
Bassi et al., 2022 \cite{MF2022}  & 3x3  & P & both & \xmark & Model-free  \\ 
Yi \&  Verbič, 2022 \cite{YI2022108465}  & 1x1 &  & export & \cmark  & Convex relaxation, explicit uncertainty \\ 
Ochoa et al., 2022 \cite{ochoa2022reactive}  &  3x3 & P+Q & both & \cmark & Conceptual \\ %
Gerdroodbari et al., 2022\cite{ZABIHINIAGERDROODBARI2022119757}  & 4x4 & P+Q & both & \xmark & Iterative using PF solver  \\ %
Milford \& Krause, 2021 \cite{9715663}  & 4x4 & P & both & \cmark   & Linearized in state estimation solution  \\ 
Project EDGE, 2022 \cite{edge2022} & (?) & P+Q & both & \xmark & Iterative using PF solver/heuristics \\ %
Antić et al. & 3x3 & P+Q & export & \cmark & Exact, nonlinear, nonconvex \\
\hline
\end{tabular}
}
\end{table*}

Exact optimization models for power networks with variables for complex power are inevitably nonlinear and nonconvex. 
Due to perceived scalability issues, many authors shy away from developing nonlinear programming models, and instead develop a linearized model of the physics to pass on to linear programming solvers \cite{9248975,Liu2022,9816082,9692618,YI2022108465,9715663}.

Neglecting phase unbalance leads to very optimistic estimates of import and export limits, that will not be achievable in the field. 
Linearized power flow formulations such as \textsc{LinDist3Flow} \cite{claeysdistflow} only have variables for voltage magnitude and do not include voltage unbalance factor (VUF) limitations. 

Nonlinear models are nevertheless capable of representing all the limits without approximation. 
Neglecting network losses, e.g. in \textsc{LinDist3Flow}, leads to voltage magnitude accuracy problems \cite{claeysdistflow}, as the power needed to supply the losses does not cause an additional voltage drop due to the linearization. 
Linearization can help scaling up optimization models that consider uncertainty though, which has recently been explored in \cite{Liu2022,YI2022108465}.

We conclude that there is a gap in the literature on optimization-based DOE calculation using exact \emph{nonlinear} models of the \emph{unbalanced} distribution network. 
Papers frequently also ignore voltage unbalance metrics such as VUF. 
To overcome the identified research gap, following contributions are proposed:
\begin{itemize}
    \item In this paper, we explore the impact of different technical constraints on the values calculated for DOEs, including a transformer's and lines' current and voltage magnitude and unbalance constraints
    \item The exact nonlinear physics without approximation, with an unbalanced model of an LV network is used in the DOEs calculation. 
    \item We illustrate the impact on two case studies of real-world network datasets. 
\end{itemize}
Note that \emph{we do not design DOE calculation procedures in this work}, but instead focus on exploring the application of technical limits that influence import/export values derived by such approaches in a LV network context.

The rest of the paper is organized as follows: An exact mathematical model used in the calculation of DOEs is presented in Section \ref{sec:math_for}. Case studies with the descriptions of a Croatian and Australian LV network used in calculations are defined in Section \ref{sec:cs}, while definition of scenarios and the results of calculation are shown in Section \ref{sec:scen_res}. Finally the conclusions are given in Section \ref{sec:conc}.

\section{Mathematical formulation}\label{sec:math_for}
The calculation of the export DOEs presented in this paper is based on the non-convex three-phase current-voltage OPF formulation, without relaxation or linearization using the modification of \textit{pp OPF}, the tool presented in \cite{ppopf}.

Calculation of the real and imaginary part of the voltage drop across all phases $p$ for a branch $l$ connecting buses $i$ and $j$ can be represented with eqs. (\ref{eq:voltage_drop_re}) and (\ref{eq:voltage_drop_im}). 
\begin{equation}
    U_{j,p,t}^{re} = U_{i,p,t}^{re} - \sum_{q \in \{a,b,c\}}R_{l,pq}\cdot I_{l,ij,q,t}^{re} + \sum_{q \in \{a,b,c\}}X_{l,pq}\cdot I_{l,ij,q,t}^{im}   
\label{eq:voltage_drop_re}
\end{equation}
\begin{equation}
    U_{j,p,t}^{im} = U_{i,p,t}^{im} - \sum_{q \in \{a,b,c\}}R_{l,pq}\cdot I_{l,ij,q,t}^{im} - \sum_{q \in \{a,b,c\}}X_{l,pq}\cdot I_{l,ij,q,t}^{re}  
\label{eq:voltage_drop_im}
\end{equation}

Active and reactive power in branch $l$ flow from node $i$ to node $j$ are constrained with eqs. (\ref{eq:line_act_power_calc_phase}) and (\ref{eq:line_react_power_calc_phase}).
\begin{equation}
    P_{l,ij,p,t} = U_{i,p,t}^{re}\cdot I_{l,ij,p,t}^{re} + U_{i,p,t}^{im}\cdot I_{l,ij,p,t}^{im} 
    \quad \forall (l,i,j)
    \label{eq:line_act_power_calc_phase}
\end{equation}
\begin{equation}
    Q_{l,ij,p,t} = U_{i,p,t}^{im}\cdot I_{l,ij,p,t}^{re} - U_{i,p,t}^{re}\cdot I_{l,ij,p,t}^{im} \quad \forall (l,i,j)
    \label{eq:line_react_power_calc_phase}
\end{equation}

Since the presented model is defined as a current-voltage formulation, it is necessary to calculate real and imaginary part of currents caused by nodal demand $d$ and generation $g$, as shown in eqs. (\ref{eq:p_element}) and (\ref{eq:q_element}).
\begin{equation}
    P_{g/d,p,t} = U_{i,p,t}^{re} \cdot I_{g/d,p,t}^{re} + U_{i,p,t}^{im} \cdot I_{g/d,p,t}^{im} \quad \forall (g/d,i)
\label{eq:p_element}
\end{equation}
\begin{equation}
   Q_{g/d,p,t} = U_{i,p,t}^{im} \cdot I_{g/d,p,t}^{re} - U_{i,p,t}^{re} \cdot I_{g/d,p,t}^{im} 
\label{eq:q_element}
\end{equation}

Finally, in order to ensure Kirchoff's Current Law (KCL), for both real and imaginary part, eq. (\ref{eq:kirchoff_current}) is introduced.
\begin{equation}
    I_{d,i,p,t }^{re/im} - I_{g,i,p,t }^{re/im} - I_{l,h \rightarrow i,p,t}^{re/im} + I_{l,i \rightarrow j,p,t}^{re/im} = 0
    \label{eq:kirchoff_current}
\end{equation}

Eq. \eqref{eq:max_current} is the expression for constraining the value of the line $l$ current, while eq. (\ref{eq:node_volt_limit}) and eqs. (\ref{eq:vuf})-(\ref{eq:volt_pos_square_mag}) are the common expressions used for constraining the values of voltage magnitude and voltage unbalance factor. Maximum values of current are defined as parameters of input data, i.e., each line in both the Croatian and Australian case study has the predefined fixed value of the allowed current. In both Australian and Croatian low voltage networks, the maximum value of voltage $U^{max}$ is equal to 1.1 p.u. Minimum voltage $U^{min}$ in the Australian LV grid is equal to 0.94 p.u., while in the Croatian case, it is equal to 0.9 p.u. Maximum values of $VUF$ are equal to 2\%, which is the threshold defined in national grid codes and other relevant standards.

\begin{equation}
    (I_{l,ij,p,t}^{re})^2 + (I_{l,ij,p,t}^{im})^2 \leq (I_{l,ij}^{max})^2
    \label{eq:max_current}
\end{equation}

\begin{equation}
    (U^{min})^2 \leq (U_{i,p,t}^{re})^2 + (U_{i,p,t}^{im})^2 \leq (U^{max})^2
    \label{eq:node_volt_limit}
\end{equation}
\begin{equation}
    \frac{|U_{2,n}|^2}{|U_{1,n}|^2} \leq (VUF_n^{max})^2
    \label{eq:vuf}
\end{equation}

\begin{equation}
\begin{gathered}
    |U_{2,n}|^2=[U_{a,n}^{re}-\frac{1}{2}\cdot (U_{b,n}^{re} + U_{c,n}^{re})+\frac{\sqrt{3}}{2}\cdot (U_{b,n}^{im} - U_{c,n}^{im})]^2 + \\
    [U_{a,n}^{im}-\frac{1}{2}\cdot (U_{b,n}^{im} + U_{c,n}^{im})-\frac{\sqrt{3}}{2}\cdot (U_{b,n}^{re} - U_{c,n}^{re})]^2
\end{gathered}
\label{eq:volt_neg_square_mag}
\end{equation}

\begin{equation}
\begin{gathered}
    |U_{1,n}|^2=[U_{a,n}^{re}-\frac{1}{2}\cdot (U_{b,n}^{re} + U_{c,n}^{re})-\frac{\sqrt{3}}{2}\cdot (U_{b,n}^{im} - U_{c,n}^{im})]^2 + \\
    [U_{a,n}^{im}-\frac{1}{2}\cdot (U_{b,n}^{im} + U_{c,n}^{im})+\frac{\sqrt{3}}{2}\cdot (U_{b,n}^{re} - U_{c,n}^{re})]^2
\end{gathered}
\label{eq:volt_pos_square_mag}
\end{equation}

In the first case, the only objective is to maximize the sum of active power of all generators $g$ connected to a network over time periods $t$, as shown in eq. (\ref{eq:obj_func_1}). Reactive power is defined as a variable but its value is not observed within the objective function.

\begin{equation}
    \max \sum_{t \in T} \sum_{i \in N_g} P_{g,p,t}
    \label{eq:obj_func_1}
\end{equation}
This objective is to be interpreted as the maximum simultaneous export of DGs, with the optimized maximum values defining the export limits for the customers. 
In reality, distribution utilities may deviate from this objective to make the assignment of export capacity more fair. 

In the second case, we only maximize the reactive  power margins. 
We introduce two variables to represent the import/export of reactive power separately:
\begin{equation}
     Q_{g,p,t} = Q^{+}_{g,p,t} -  Q^{-}_{g,p,t}, \quad  Q^{+}_{g,p,t}, Q^{-}_{g,p,t} \geq 0
     \label{eq:q_aux_1}
\end{equation}
We introduce an auxiliary variable $Q^{aux}_{g,p,t}$. 
\begin{equation}
     Q^{aux}_{g,p,t} \leq Q^{-}_{g,p,t}, \quad
      Q^{aux}_{g,p,t} \leq Q^{+}_{,p,t}, 
      \quad Q^{aux}_{g,p,t} \geq 0
      \label{eq:q_aux_2}
\end{equation}
And define the objective with eq. \eqref{eq:obj_func_2}, 
\begin{equation}
    \max \sum_{t \in T} \sum_{i \in N_g} Q^{aux}_{g,p,t}.
    \label{eq:obj_func_2}
\end{equation}

\section{Case studies}\label{sec:cs}
In order to exclude the results and conclusions that are tested on synthetic, benchmark networks or are valid for only one certain case, case studies are defined for Croatian and Australian LV networks, that are characterized with different layouts, technical parameters of network elements, and end-users' consumption. Defined case studies used in simulations and the calculation of DOEs are:

\begin{itemize}
    \item Case study 1 - Croatian LV network
    \begin{itemize}
        \item[1a)] Objective function is defined with eq.  \eqref{eq:obj_func_1}
        \item[1b)]  Objective function is defined with eqs. \eqref{eq:q_aux_1}-\eqref{eq:obj_func_2}
    \end{itemize}
    \item Case study 2 - Australian LV network
    \begin{itemize}
        \item[2a)] Objective function is defined with eq.  \eqref{eq:obj_func_1}
        \item[2b)]  Objective function is defined with eqs. \eqref{eq:q_aux_1}-\eqref{eq:obj_func_2}
    \end{itemize}
\end{itemize}

In our case studies we consider the load to be fixed and unsheddable. Therefore we focus on maximizing export instead.

\subsection{Croatian LV Network}
Croatian LV networks are 3-phase 4-wire networks and depending on the part of Croatia, they are mostly constructed with underground cables in urban parts and with overhead lines in rural parts of the country. An LV network used in simulations is an urban network with 65 nodes, an MV/LV transformer, and 63 cables, defined with their positive and zero sequence resistance and reactance, which are then used in calculating the three-phase impedance matrix of a network.

In a network defined in the Croatian case study, there are 43 three-phase connected end-users, and their phase consumption curves are created from the measurements collected from smart meters. In the case study in this paper, all distributed generators (DGs) are observed as single-phase with a randomly selected connection phase in each node. 

\subsection{Australian LV Network}
We choose network `N' from the public data release of the CSIRO LV feeder taxonomy project \cite{TaxonomyStudy}, and use the improved impedance data proposed in \cite{anson2022}. 
This Australian network is a 3-phase 4-wire  with a multigrounded neutral overhead backbone, with aerial bundled conductors for the service line (between the feeder and the house). 
Kron's reduction of the neutral has been performed throughout. A modeled Australian LV network consists of 99 LV nodes and 98 cables.

End-users are connected to 63 nodes, and 62 of them are single-phase connected to a network, while only one is connected three-phase. Same as in the Croatian case study, the consumption curve for each end-user is created from real-world smart meter data. In the Australian case study, DGs are single-phase connected, to the same phase as end-users.

\section{Scenarios and results}\label{sec:scen_res}
The main goal of the presented idea is to show the importance of different technical constraints in calculating DOEs in LV distribution networks. By defining different scenarios, it is possible to assess the value of each technical constraint and their contribution to calculating export DOEs in LV networks but also to identify the shortcoming in already existing approaches that neglect constraints such as current or unbalance constraints. Five defined scenarios include different sets of technical limits:

\begin{enumerate}
    \item DG production is constrained by the limitations defined by the Croatian (3.68 kVA) and Australian grid code (5 kVA/phase)
    \item Voltage magnitude and unbalance constraints are considered while transformer's and lines' current constraints are neglected
    \item Voltage and current grid constraints are considered while the voltage unbalance constraint is neglected
    \item Current magnitude and voltage unbalance constraints are considered while the voltage magnitude constraint is neglected
    \item All constraints are considered (voltage and current magnitude, voltage unbalance)
\end{enumerate}

Scenario 1 is defined to compare the possibility of installation of DGs according to current standards which need to be satisfied for system operators to allow the installation. Scenario 2 is chosen to determine the error of models that neglect current constraints. Most of the papers neglect the voltage unbalance constraint and the authors in general do not consider its value in the calculation of DOEs. To overcome this shortcoming, we identified Scenario 3. Even though the voltage magnitude constraints are always considered in already developed models, it is hard to assess its value compared to other technical constraints. Therefore, we define Scenario 4. Finally, Scenario 5 is the one in which all constraints are considered to use a more precise model. Defining multiple scenarios allows the assessment of the value of different network constraints in the context of calculating dynamic operating envelopes.

Table \ref{tab:production} is given as a brief summary of the DOEs calculation for both Croatian and Australian case studies. Active production in cases 1a) and 2a) emphasizes the need for implementing DOEs and abandoning traditional approaches in which production power is limited without considering the network conditions. In scenarios 3-5, daily production is larger by more than 2 MWh compared to Scenario 1, which is limited by the national grid codes. The production is even larger in Scenario 2 in which the current constraint is neglected. Due to the increase in values of active and reactive power, the such approach brings values of voltage magnitude close to the bounds but also increases the current flow which is multiple times higher than any feasible, possible solution. These results present a warning for models that rely on a correlation between the values of nodal voltages and power instead of on a detailed model of a network. 

The change of the objective function in cases 1b) and 2b) causes a decrease in the daily production in both the Croatian and Australian case studies. The used modeling approach gives values of reactive power production that are close to zero. With a such set of values of DGs' reactive power, daily active production is between 3 and 4 MWh lower than in cases 1a) and 2a). Active production in scenarios 3-5 is even smaller than the potential production constrained only by the national grid codes. In the Croatian case study, the value of production in Scenario 2 in which the current constraint is neglected is more than 1 MWh smaller than in Scenario 1, which only shows the importance of unbalance constraint in some network topologies and highlight the flaws of approaches that neglect it.

\begin{table}[htbp]
\centering
\caption{Potential daily production of DGs for different scenarios}
\label{tab:production}
\begin{tabular}{crlrc}
\hline
           & \multicolumn{4}{c}{Total production (kWh)}                                                              \\ \hline
           & \multicolumn{1}{c}{1a)} & \multicolumn{1}{c}{1b)}  & \multicolumn{1}{c}{2a)} & 2b)                      \\ \hline
Scenario 1 & 3 797.76                & 3 797.76                 & 7560.00                 & 7 560.00                 \\
Scenario 2 & 63 348.53               & 2 578.36                 & 123 219.27              & 23 262.20                \\
Scenario 3 & 6 230.82                & 2 343.95                 & 9 578.21                & 5 859.42                 \\
Scenario 4 & 6 623.02                & 3 522.97                 & 9 654.57                & 5 794.44                 \\
Scenario 5 & 6 108.92                & 2 380.36                 & 9 592.73                & 5 833.06                 \\ \hline
           & \multicolumn{4}{c}{Total production (kVArh)}                                                            \\ \hline
           & \multicolumn{1}{c}{1a)} & \multicolumn{1}{c}{1b)}  & \multicolumn{1}{c}{2a)} & 2b)                      \\ \hline
Scenario 1 & 0.00                    & \multicolumn{1}{r}{0.00} & 0.00                    & \multicolumn{1}{r}{0.00} \\
Scenario 2 & -16 506.63              & \multicolumn{1}{r}{0.00} & -71 650.03              & \multicolumn{1}{r}{0.00} \\
Scenario 3 & 627.76                  & \multicolumn{1}{r}{0.00} & 879.63                  & \multicolumn{1}{r}{0.00} \\
Scenario 4 & 743.01                  & \multicolumn{1}{r}{0.00} & 1 054.44                & \multicolumn{1}{r}{0.00} \\
Scenario 5 & 660.04                  & \multicolumn{1}{r}{0.00} & 940.27                  & \multicolumn{1}{r}{0.00} \\ \hline
\end{tabular}
\end{table}

\subsection{Case Study 1}
\begin{figure}[htbp]
    \centering
   \begin{subfigure}[b]{\columnwidth}
         \centering
        \includegraphics[width=\columnwidth]{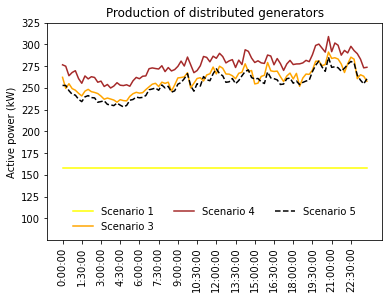}
        \caption{}
        \label{}
    \end{subfigure}
    \begin{subfigure}[b]{\columnwidth}
         \centering
        \includegraphics[width=\columnwidth]{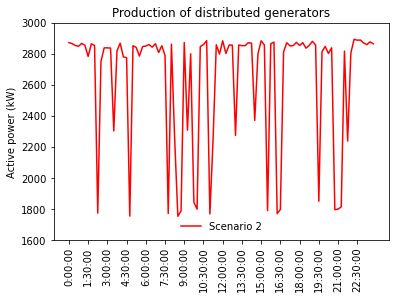}
        \caption{}
    \end{subfigure}
    \caption{Calculated export DOEs - Case Study 1a}
    \label{fig:production_cro_1a}
\end{figure}

Fig. \ref{fig:production_cro_1a} shows the results of the calculation in case 1a), when the objective function is to maximize the sum of the active power of all generators. The importance and necessity of considering different technical constraints in case 1a) can be seen in Fig. \ref{fig:production_cro_1a}  which shows the change of production in each time period for each scenario. The largest production power of DGs is in Scenario 2, the one in which current constraints are neglected, similar to model-free approaches. However, a significant decrease in production power in other scenarios clearly shows the shortcoming of such approaches. Additionally, commonly used models that neglect voltage unbalance are inadequate since the additional consideration of unbalance (Scenario 5) further limits the production of DGs compared to the scenario in which only this constraint is neglected (Scenario 3). The results in Scenario 5 clearly show the need of abandoning the concept in which the limitation of DGs connection power is constant since based on a network's layout and end-users' consumption production curve changes and is always larger than defined limitation.

\begin{figure}[htbp]
    \centering
    \includegraphics[width=\columnwidth]{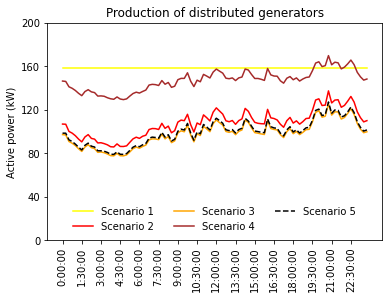}
    \caption{Calculated export DOEs - Case Study 1b}
    \label{fig:production_cro_pq}
\end{figure}

Fig. \ref{fig:production_cro_pq} shows the results for the Croatian case study 1b) when the objective function is the maximization of the reactive power margins. Defining additional constraints in the mathematical model used in case 1b) limits the active production in the observed network and for most scenarios and time periods even decreases it below the limitation defined by the Croatian grid code (Scenario 1). 

Furthermore, the results in the graph additionally emphasize the need of modeling all technical constraints as in Scenario 5, since they lead to a decrease in active production. Production in scenarios 3, 4, and 5 are lower than in case 1a) but with a similar shape of the DOE curve, unlike production in Scenario 2 which not only significantly decreased but the shape of the curve changed and the drops that previously existed do not occur.

\subsection{Case Study 2}
Fig. \ref{fig:production_au_2a} shows the results of the DOEs calculation for the Australian case study 2a). The conclusions drawn are similar to those in the Croatian one, with the exception of an even larger decrease in DGs' production power in scenarios in which the current constraint is considered. Also, the range between the upper and lower bound of production power in scenarios 3-5 is larger in the Australian case study which can be correlated with a difference in consumption profiles but also the difference in the network's layout. Another significant importance compared to a Croatian case study is the smaller impact of modeling different constraints on the optimal solution of the presented problems, with the same reasons as for the larger range of active production power values.

\begin{figure}[htbp]
    \centering
    \begin{subfigure}[b]{\columnwidth}
         \centering
        \includegraphics[width=\columnwidth]{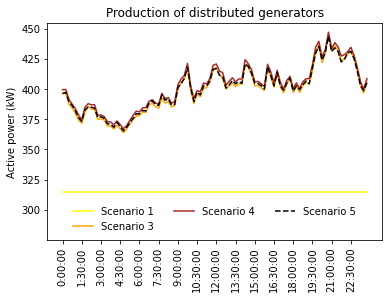}
        \caption{}
     \end{subfigure}
\end{figure}

\begin{figure}[htbp]
\ContinuedFloat
    \begin{subfigure}[b]{\columnwidth}
         \centering
    \includegraphics[width=\columnwidth]{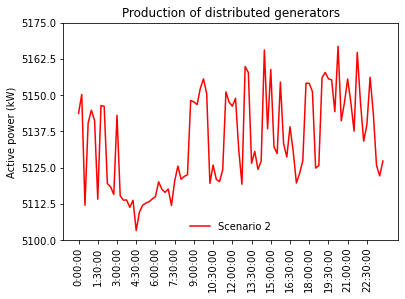}
    \caption{}
    \end{subfigure}
    \caption{Calculated export DOEs - Case Study 2a}
    \label{fig:production_au_2a}
\end{figure}

The calculation of DOEs in case study 2b) is shown in Fig. \ref{fig:production_au_2b}. According to the results, the change of the objective functions leads to limiting the active production of DGs below the value defined by the Australian grid code. Also same as in the first Australian case 2a), the impact of different technical constraints on values in the graph is not as significant as in Croatian cases. Active production in Scenario 2 is multiple times higher than production in other scenarios, the same as in the Australian case 2a). An interesting point is that in the Croatian case 1b), in which the objective function is the same as in case 2b), this difference does not exist meaning that the threshold value of voltage unbalance is reached before the value of active power increases enough to lead to high currents in an observed grid. This emphasizes the need for precise modeling but also shows a difference caused by observing two network topologies with non-similar characteristics.

\begin{figure}[htbp]
    \centering
    \begin{subfigure}[b]{\columnwidth}
         \centering
        \includegraphics[width=\columnwidth]{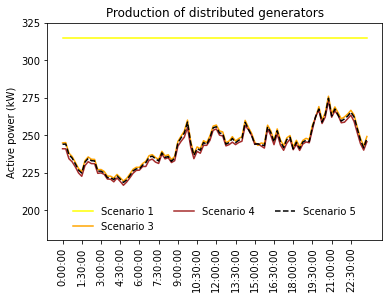}
        \caption{}
    \end{subfigure}
\end{figure}

\begin{figure}[htbp]
\ContinuedFloat
    \begin{subfigure}[b]{\columnwidth}
         \centering
        \includegraphics[width=\columnwidth]{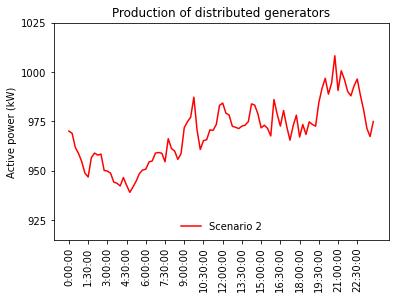}
        \caption{}
    \end{subfigure}
    \caption{Calculated export DOEs - Case Study 2b}
    \label{fig:production_au_2b}
\end{figure}

\section{Conclusions}\label{sec:conc}
This paper explored the application of different network limits in the context of potential dynamic operating envelope calculation procedures.

It is shown that the maximum export values depend on the sets of limits represented in a significant manner. 
An analysis of a broader set of networks would be interesting to understand if the different limits become binding in a different order depending on the network design. 
In certain regions, LV networks have a very sparse grounding of the neutral, e.g. only at the substation. In this case, the inclusion of neutral-to-ground voltage limits may be crucial in determining the DOE. 
Finally, in the real world, DOE calculation should be done frequently and based on a state estimate, so that congestions are avoided with high likelihood while simultaneously avoiding unnecessary curtailment of PV. The tradeoffs between the frequency of the updates, accuracy of the estimates, and interactions between customers due to congestions are to be researched further. 

Future work  includes looking at different trade-offs between customer equity and export capacity. 

\bibliographystyle{IEEEtran}
\bibliography{library}

\begin{thebibliography}{10}
\providecommand{\url}[1]{#1}
\csname url@samestyle\endcsname
\providecommand{\newblock}{\relax}
\providecommand{\bibinfo}[2]{#2}
\providecommand{\BIBentrySTDinterwordspacing}{\spaceskip=0pt\relax}
\providecommand{\BIBentryALTinterwordstretchfactor}{4}
\providecommand{\BIBentryALTinterwordspacing}{\spaceskip=\fontdimen2\font plus
\BIBentryALTinterwordstretchfactor\fontdimen3\font minus
  \fontdimen4\font\relax}
\providecommand{\BIBforeignlanguage}[2]{{%
\expandafter\ifx\csname l@#1\endcsname\relax
\typeout{** WARNING: IEEEtran.bst: No hyphenation pattern has been}%
\typeout{** loaded for the language `#1'. Using the pattern for}%
\typeout{** the default language instead.}%
\else
\language=\csname l@#1\endcsname
\fi
#2}}
\providecommand{\BIBdecl}{\relax}
\BIBdecl

\bibitem{fti2022}
\BIBentryALTinterwordspacing
``Dynamic operating envelope policy in the national electricity market,'' FTI
  Consulting, for the Australian Energy Regulator, Technical Report, 06 2022.
  [Online]. Available:
  \url{https://www.aer.gov.au/networks-pipelines/guidelines-schemes-models-reviews/review-of-regulatory-framework-for-flexible-export-limit-implementation}
\BIBentrySTDinterwordspacing

\bibitem{CM2022}
\BIBentryALTinterwordspacing
``{Review of Dynamic Operating Envelopes Adoption by DNSPs},'' CutlerMerz,
  Technical Report, 2022. [Online]. Available:
  \url{https://arena.gov.au/assets/2022/07/review-of-dynamic-operating-envelopes-from-dnsps.pdf}
\BIBentrySTDinterwordspacing

\bibitem{aer2022}
\BIBentryALTinterwordspacing
``Flexible export limits issues paper,'' Australian Energy Regulator, Technical
  Report, 010 2022. [Online]. Available:
  \url{https://www.aer.gov.au/networks-pipelines/guidelines-schemes-models-reviews/review-of-regulatory-framework-for-flexible-export-limit-implementation}
\BIBentrySTDinterwordspacing

\bibitem{FOBES2020106664}
D.~M. Fobes, S.~Claeys, F.~Geth, and C.~Coffrin,
  ``{PowerModelsDistribution.jl}: An open-source framework for exploring
  distribution power flow formulations,'' \emph{Elect. Power Syst. Res.}, vol.
  189, p. 106664, 2020.

\bibitem{ppopf}
\BIBentryALTinterwordspacing
T.~Antić, A.~Keane, and T.~Capuder, ``Pp opf -- pandapower implementation of
  three-phase optimal power flow model,'' 2022. [Online]. Available:
  \url{https://arxiv.org/abs/2211.11032}
\BIBentrySTDinterwordspacing

\bibitem{ZABIHINIAGERDROODBARI2022119757}
Y.~{Zabihinia Gerdroodbari}, M.~Khorasany, and R.~Razzaghi, ``Dynamic {PQ}
  operating envelopes for prosumers in distribution networks,'' \emph{Applied
  Energy}, vol. 325, p. 119757, 2022.

\bibitem{edge2022}
\BIBentryALTinterwordspacing
``{Lessons Learnt 2},'' Project EDGE, Technical Report, 2022. [Online].
  Available:
  \url{https://arena.gov.au/assets/2022/11/project-edge-lessons-learnt-2.pdf}
\BIBentrySTDinterwordspacing

\bibitem{EVOLVE2020}
\BIBentryALTinterwordspacing
L.~Blackhall, ``On the calculation and use of dynamic operating envelopes,''
  Project EVOLVE, Technical Report, 2020. [Online]. Available:
  \url{https://arena.gov.au/assets/2020/09/on-the-calculation-and-use-of-dynamic-operating-envelopes.pdf}
\BIBentrySTDinterwordspacing

\bibitem{9248975}
K.~Petrou, M.~Z. Liu, A.~T. Procopiou, L.~F. Ochoa, J.~Theunissen, and
  J.~Harding, ``Operating envelopes for prosumers in {LV} networks: A weighted
  proportional fairness approach,'' in \emph{IEEE PES Innovative Smart Grid
  Techn. Conf. Europe}, 2020, pp. 579--583.

\bibitem{Liu2022}
B.~Liu and J.~H. Braslavsky, ``Robust operating envelopes for der integration
  in unbalanced distribution networks,'' \emph{[math.OC]}, 2022.

\bibitem{9816082}
M.~Z. Liu, L.~F. Ochoa, P.~K.~C. Wong, and J.~Theunissen, ``{Using OPF-Based
  Operating Envelopes to Facilitate Residential DER Services},'' \emph{IEEE
  Trans. Smart Grid}, vol.~13, no.~6, pp. 4494--4504, 2022.

\bibitem{9692618}
M.~Z. Liu, L.~F. Ochoa, T.~Ting, and J.~Theunissen, ``Bottom-up services \&
  network integrity: The need for operating envelopes,'' in \emph{26th Int.
  Conf. Exhib. Elect. Distrib.}, vol. 2021, 2021, pp. 1944--1948.

\bibitem{MF2022}
V.~Bassi, D.~Jaglal, L.~Ochoa, T.~Alpcan, and C.~Leckie, ``{Deliverables 1-2-3a
  Model-Free Voltage Calculations and Operating Envelopes},'' University of
  Melbourne, Technical Report, 7 2022.

\bibitem{YI2022108465}
Y.~Yi and G.~Verbič, ``Fair operating envelopes under uncertainty using chance
  constrained optimal power flow,'' \emph{Elect. Power Syst. Res.}, vol. 213,
  p. 108465, 2022.

\bibitem{ochoa2022reactive}
\BIBentryALTinterwordspacing
L.~Ochoa, M.~Liu, J.~Theunissen, and N.~Regan, ``Reactive power and voltage
  regulation devices to enhance operating envelopes,'' Technical Report, 06
  2022. [Online]. Available:
  \url{https://www.researchgate.net/publication/361405704\_Reactive\_Power\_and\_Voltag\_Regulation\_Devices\_to\_Enhance\_Operating\_Envelopes}
\BIBentrySTDinterwordspacing

\bibitem{9715663}
T.~Milford and O.~Krause, ``Managing {DER} in distribution networks using state
  estimation \& dynamic operating envelopes,'' in \emph{IEEE PES Innovative
  Smart Grid Tech. Conf. Asia}, 2021, pp. 1--5.

\bibitem{claeysdistflow}
S.~Claeys, F.~Geth, M.~Sankur, and G.~Deconinck, ``No-load linearization of the
  lifted multi-phase branch flow model: Equivalence and case studies,'' in
  \emph{IEEE PES Innovative Smart Grid Techn. Conf. Europe}, 2021, pp. 1--5.

\bibitem{TaxonomyStudy}
\BIBentryALTinterwordspacing
{F. Geth, et al.}, ``{National Low-Voltage Feeder Taxonomy Study.}'' CSIRO
  Energy, Technical Report, 2021. [Online]. Available:
  \url{https://doi.org/10.25919/2tas-7213}
\BIBentrySTDinterwordspacing

\bibitem{anson2022}
C.~H. Tam, F.~Geth, and M.~Nadarajah, ``An inclusive model for a practical
  low-voltage feeder with explicit multi-grounded neutral wire,'' in \emph{IEEE
  Sust. Power Energy Conf.}, 2022.

\end{thebibliography}
\end{document}